\documentclass[pra, aps, twocolumn, groupaddress, superscriptaddress]{revtex4}

\usepackage{slashed}
\usepackage{graphicx}
\usepackage{subfigure}
\usepackage[usenames, dvipsnames]{color}
\usepackage{graphics}
\usepackage{hyperref}
\usepackage{bm}
\usepackage{amsfonts}
\hypersetup{backref,
colorlinks=true,
linkcolor=blue,
linktoc=page,
citecolor=blue}

\everymath{\displaystyle} 
 
\newcommand{\pc}[1]{\ensuremath{\left(#1\right)}}

\begin{document}

\title{Phonon-polaritons in Bose-Einstein condensates induced by Casimir-Polder interaction with graphene}

\author{H. Ter\c cas}

\email{hugo.tercas@uibk.ac.at}

\affiliation{Institute for Theoretical Physics, University of Innsbruck, 6020 Innsbruck, Austria}
\affiliation{Institute for Quantum Optics and Quantum Information,\\ Austrian Academy of Sciences, 6020 Innsbruck, Austria}

\author{S. Ribeiro}
\affiliation{School of Physics and Astronomy, University of Southampton, Southampton, SO17 1BJ, United Kingdom}

\author{J. T. Mendon\c{c}a}

\affiliation{Instituto de F\'isica, Universidade de S\~ao Paulo, S\~ao Paulo SP, 05508-090 Brasil}
\affiliation{IPFN, Instituto Superior T\'ecnico, 1049-001 Lisboa, Portugal}

\begin{abstract}

We consider the mechanical coupling between a two-dimensional Bose-Einstein condensate with a graphene sheet via the vacuum fluctuations of the electromagnetic field which are at the origin of the so-called Casimir-Polder potential. By deriving a self-consistent set of equations governing the dynamics of the condensate and the flexural (out-of-plane) modes of the graphene, we can show the formation of a new type of purely acoustic quasi-particle excitation, a phonon-polariton resulting from the coherent superposition of quanta of flexural and Bogoliubov modes.

\end{abstract}
\maketitle

\section{Introduction}

The Casimir effect is a consequence of the field-theoretical description of the quantum vacuum and results directly from the quantization of the electromagnetic field. Traditionally derived for two infinite uncharged metallic plates placed only a few nanometers apart \cite{casimir, genet}, this quantum-mechanical effect has also been investigated in the context of atoms interacting with surfaces \cite{babb, milton}. Casimir-Polder (CP) forces have been a subject of research of its own, both in view of nano-technological applications \cite{chan} and motivated by the possibility of probing fundamental forces at the submicron scale \cite{dimopoulos}. Indeed, the effects of a dispersive potential due to a macroscopic surface on an atom, both at zero and finite temperature, are already quite well established \cite{StefanB_Book}.\par

Moreover, experiments with Bose-Einstein condensates (BEC) near a surface have attracted a special attention since the early days of microtraps \cite{lin, leanhardt, harber}: if, in one hand, the understanding of the influence of vacuum forces is crucial to the operations in atomic microtraps, on the other, the measurements based on quantum optics of ultracold gases are very accurate, making them a very desirable candidate to probe vacuum forces \cite{antezza}.  In fact, the influence of the solid-state substrate on the atomic dynamics of a cold gases has been the target of several experiments. As it has been shown, an atomic cloud near a rough surface (typical distances of 1~$\mu$m)  may undergo a matter-wave Anderson localization in a random potential \cite{PRL105_210401_2010}. Also, by rotating a corrugated plate separated of a few microns from a BEC, the nucleation of quantized vortices is theoretically predicted \cite{EPL92_40010_2010}.  
More recently, Bender et al. have used BEC to harvest information about the shape of the potential landscape of a solid grating, in an excellent agreement with the theoretical predictions \cite{bender}. \par
An important and recent activity with ultracold atoms includes mechanical coupling via vacuum forces. Experiments have put in evidence the resonant coupling of mechanical BEC modes to a micromechanical oscillator \cite{PRL104_143002_2010}, as well as the backaction of the atomic motion onto the membrane \cite{PRL107_223001_2011}. There are also theoretical propositions of how creating a force in a graphene sheet due to highly excited (Rydberg) atomic states \cite{PRA88_052521_2013}. Actually, the interest in the family of fullerenes has growing since the recent propositions of sympathetic cooling of carbon nanotubes, via CP interaction, by a laser cooled atomic gas \cite{weiss}. However, to the best of our knowledge, a  completed treatment of the coupling of the flexural modes of a graphene membrane and a BEC via vacuum forces has not yet been discussed. In this paper, we will provide a microscopic description of the problem. \par
\begin{figure}[t!]
\includegraphics[scale=0.6]{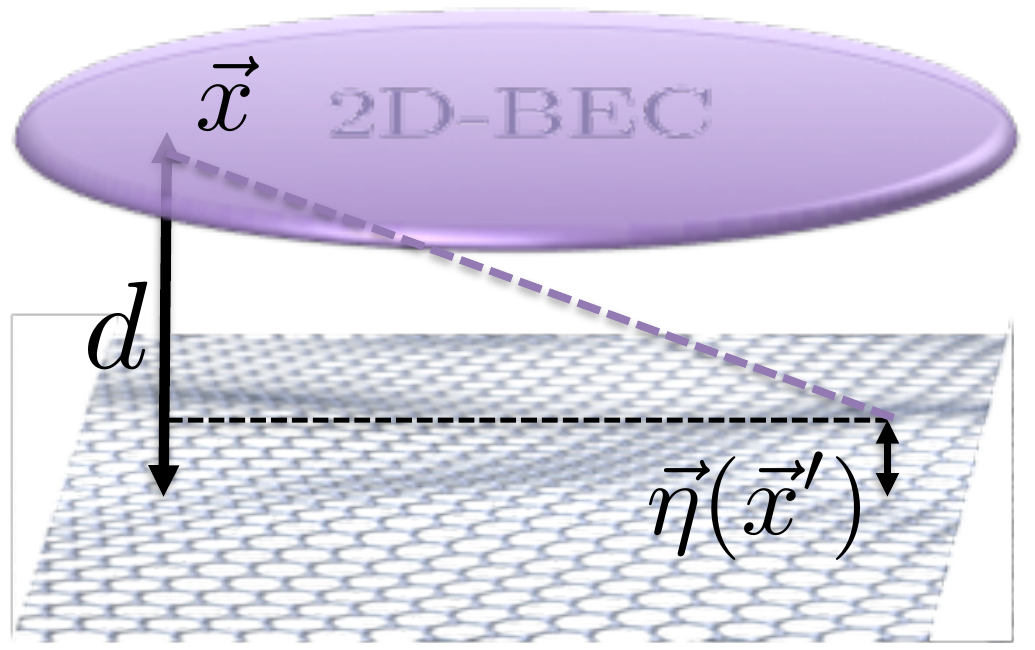}
\caption{(color online) Schematic representation of the system. A two-dimensional BEC is placed at a distance $d$ from a monolayer graphene sheet. The a ripple at the position $\mathbf{x}'$ provoques a small deformation in the Casimir-Polder potential at a position $\mathbf{x}$ in the BEC and vice-versa.}
\label{fig1}
\end{figure}
Benefiting from its remarkable mechanical and transport properties \cite{neto}, monolayer graphene is a natural candidate to perform atom-surface interfaces. Due to thermal fluctuations, the membrane may undergo mechanical out-of-plane vibrations (flexural phonons), which can be well described within the Kirchoff-plate theory of elasticity \cite{amorim}. A recent experimental study makes use of a cavity optomechanical protocol to cool down the zero-point flexural mode of a suspended graphene sheet \cite{song}. By previously cooling the graphene sheet (with the help of in a dilution refrigerator, for example), the flexural modes can be quantized. Thus, in this paper, we investigate the dynamics of a two-dimensional BEC interacting with a monolayer graphene sheet via the Casimir-Polder potential (see Fig. (\ref{fig1}) for a schematic illustration). We proceed to a full quantization of the system in order to harvest the coherent coupling between the two phonon modes, leading to a new type of polariton excitation: a purely acoustic phonon-polariton. We show that for sufficiently large separation distances $d$, heating of the condensate via vacuum fluctuations is negligible, showing that the polaritons may exist in the strong-coupling regime. In Sec. II, we present the governing equations in terms of the mean-field equations. The details of the CP potential for a $^{87}$Rb condensate are provided in Sec. III. In Sec. IV, the dispersion relation of the phonon-polariton modes and the BEC heating rate are derived. Finally, in Sec. V some concluding remarks are stated. 

\section{Governing equations: mean-field description}
 
In a general way, the potential between a single neutral atom and a surface is given by
\begin{equation}
U_S=\frac{C_\nu}{d^\nu},
\end{equation}vacuum
where $d$ is the separation distance between the atom and the surface and $C_\nu$ is the strength of the interaction, which depends on the both the atomic polarizability and on the electromagnetic properties of the surface. Let $\Psi(\mathbf{r})=\psi(z)\psi(\mathbf{x})$ be the condensate order parameter, normalized to the number of atoms such as $N=\int \vert \Psi\vert^2 d\mathbf{r}$, where $\mathbf{x}=(x,y)$ is the plane coordinate. The dynamics of the two-dimensional BEC can then be described in terms of the Gross-Pitaevskii equation
\begin{equation}
i\hbar \frac{\partial }{\partial t}\psi(\mathbf{x})=-\frac{\hbar ^2\nabla^2}{2m}\psi(\mathbf{x})+g\vert \psi(\mathbf{x}) \vert^2 \psi(\mathbf{x})+U\psi(\mathbf{x}),
\label{GP1}
\end{equation}
where $g=g_\mathrm{3D}/(2\sqrt{2\pi}\ell_z)$, with $g_\mathrm{3D}=4\pi \hbar^2 a/m$, is the effective 2D coupling constant, $a$ is the atomic scattering length and $\ell_z=(\hbar/m\omega_z)^{1/2}$ is the transverse harmonic length. The atoms in the BEC feel a vacuum-field potential of the form
\begin{equation}
U=\int \frac{d\mathbf{x}'}{\sqrt{\mathcal{V}}}   \frac{C_\nu }{\left[\left(d-{\bm \eta}(\mathbf{x}')\right)^2+\vert \mathbf{x}-\mathbf{x'}\vert^2\right]^{\nu/2}},
\label{potential1}
\end{equation}
where ${\bm \eta}(\mathbf{x})$ is a small deformation on the graphene surface (ripple) located at the position $\mathbf{x}$ and $\mathcal{V}$ is the quantization surface. On the other hand, the mechanical properties of the graphene sheet can be easily derived from the Kirchoff-Love plate theory \cite{amorim}. The Lagrangian density of the sheet can be written as $\mathcal{L}=\rho \dot {\bm \eta}^2/2-\mathcal{W}$, with $\rho$ standing for the graphene mass density, and the potential energy per unit surface can be expressed as
\begin{equation} 
\mathcal{W}=\frac{1}{2}D\left(\nabla^2{\bm \eta}\right)^2+ \gamma_x \left(\frac{\partial {\bm \eta}}{\partial x}\right)^2+\gamma_y \left(\frac{\partial {\bm \eta}}{\partial y}\right)^2+\frac{1}{2}\kappa \bm \eta^2,
\end{equation}
where $D$ is the bending stiffness, $\gamma_x$ ($\gamma_y$) is the tension along the $x$ ($y$) direction to the clamping with the substrate and $\nabla^2{\bm \eta}$ is the local curvature. The term $\kappa=d^2U/d{\bm \eta}^2\vert_{{\bm \eta}=0}$ is the restoring force in the harmonic approximation, which acts as a charge on the surface of the sheet through the CP potential,
\begin{equation}
\kappa=\nu C_\nu\int d\mathbf{x}' \vert \psi (\mathbf{x}')\vert^2 \frac{(1+\nu)d^2-\vert \mathbf{x}-\mathbf{x}'\vert^2}{\left(d^2+\vert \mathbf{x} - \mathbf{x}' \vert^2 \right)^{2+\nu/2}}.
\end{equation}
From the Euler-Lagrange equations, we can obtain the Kirchoff-Love equation for the flexural modes
\begin{equation}
\rho \frac{\partial ^2 {\bm \eta}}{\partial t^2}+D \nabla^4{\bm \eta} +\kappa{\bm \eta}=0,
\label{plate1}
\end{equation}
where we assume, for simplicity, the self-suspended case $\gamma_x=\gamma_y=0$. Eqs. (\ref{GP1}) and (\ref{plate1}) form a self-consistent set of equations for the condensate field $\psi(\mathbf{x})$ and for the deformation field ${\bm \eta}(\mathbf{x})$, being the governing equations to be quantized in Sec. IV.
\begin{figure}
\includegraphics[scale=0.26]{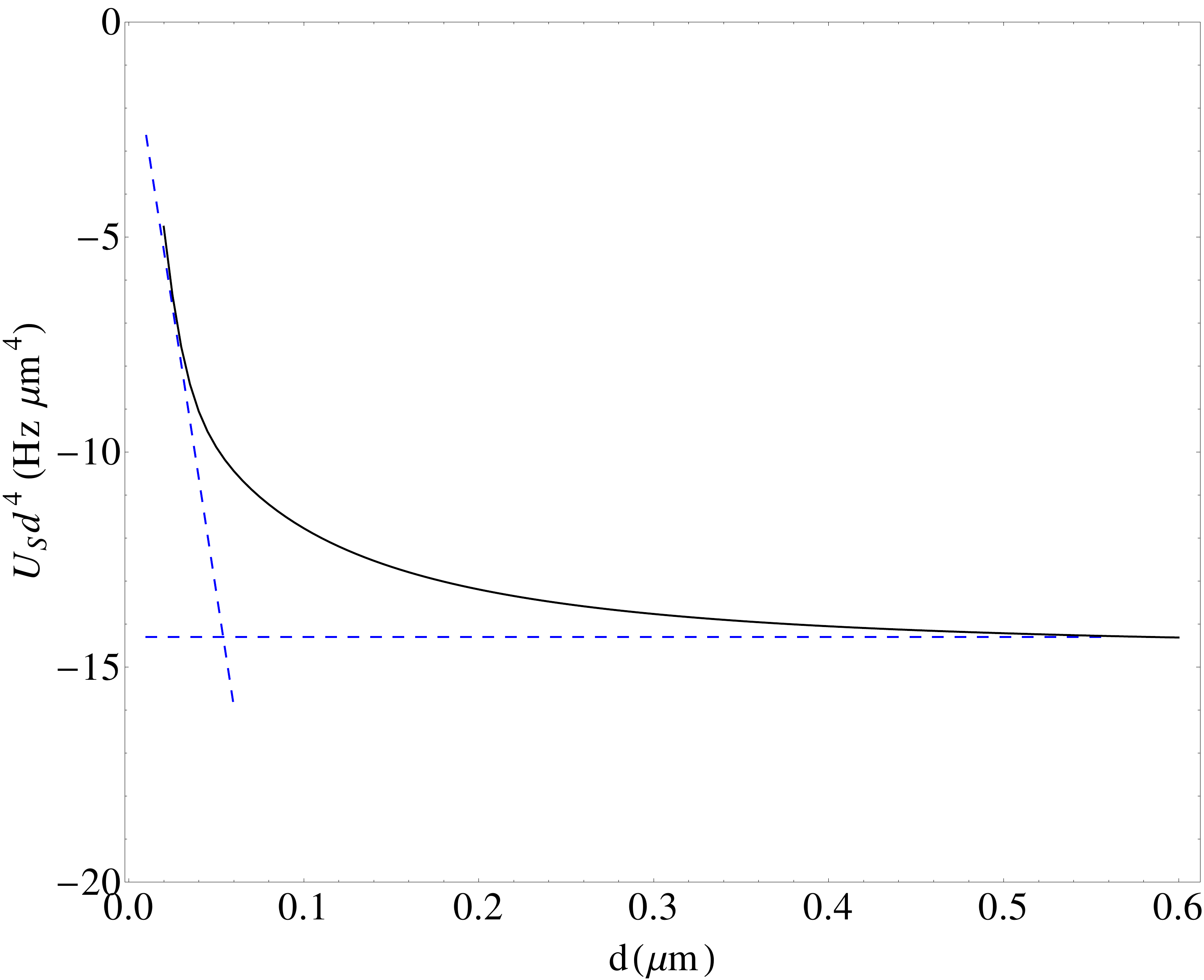}
\caption{(color online) Casimir-Polder potential of a $^{87}$Rb atom near the surface of a graphene sheet. For short distances, the potential scales as $\sim d^3$ (oblique line). The dependence on the distance as $\sim 1/d^4$ (horizontal line) is obtained for separations larger than $d=0.5~\mu$m.}
\label{fig2}
\end{figure}
\par

In the absence of coupling (or, equivalently, in the limit of very large distances $d\rightarrow \infty$), Eqs. (\ref{GP1}) and (\ref{plate1}) can be easily solved by taking the Bogoliubov prescription $\psi=e^{-i\mu t/\hbar}\left[\psi_0+\sum_k \left(u_k e^{i\mathbf{k}\cdot \mathbf{x}-i\omega t}+v_k^* e^{-i\mathbf{k}\cdot \mathbf{x}+i\omega t}\right) \right]$ and the expansion ${\bm \eta}= \sum_k {\bm \eta}_k e^{i\mathbf{k}\cdot \mathbf{x}-i\omega t} $, which respectively yield to two stable modes,
\begin{equation}
\omega_B=\left(c_s^2k^2+\frac{1}{4}c_s^2\xi^2k^4\right)^{1/2}, \quad \omega_C=\left(\beta^2 k ^4+\omega_0^2\right)^{1/2}.
\end{equation}
Here, $c_s=(g\vert \psi_0\vert^2/m)^{1/2}$ and $\xi=\hbar(g\vert \psi_0\vert^2m)^{-1/2}$ are the condensate sound speed and healing length, respectively, $\beta=(D/\rho)^{1/2}$ is the specific stiffness and $\omega_0=\sqrt{\kappa/\rho}$. In agreement with an experimentally feasible situation, we choose a $^{87}$Rb condensate with an areal density of $n_0\equiv \vert \psi_0 \vert^2\sim 10^8$ cm$^{-2}$, trapped along the $z$-direction by a harmonic trap frequency $\omega_z\sim 2\pi\times 1500$ Hz, $a=5.5$ nm, which yields $\xi\sim 0.1$ $\mu$m and $c_s\sim 0.1$ mm/s. On the other hand, for the graphene sheet we have $\rho=0.761$ mg/m$^2$ and $D\sim1.5 $ eV \cite{lambin, fasolino1, fasolino2}, leading to $\beta=6.1\times 10^{-7}$m$^2$/s. This simply means that the BEC phonons and the ripples in the graphene sheet propagate at very different frequencies. In fact, for wavevectors of the order $k\sim 10~\mu$m$^{-1}$ (typical of the BEC phonon-like excitations), $\omega_B/\omega_C\sim c_s/\beta k\simeq 10^{-3}$. Such a picture is very similar to what happens in semiconductor microcavities, where strong coupling - allowing for the formation of exciton-polaritons - occurs between two modes of very different energy scales: photons and excitons. Because of their very small effective mass, photons exhibit a parabolic dispersion relation, while the excitonic dispersion is almost flat for the relevant wavevectors \cite{guillaumebook, carusotto}. The coupling is then provided via the dipolar interaction between the photons and the excitons. In the next section, we show that the same thing happens here, where the Bogoliubov excitations in the condensate, possessing a very flat dispersion, are the analogue of the semiconductor excitons, while the mechanical vibrations of the graphene sheet work as massive photons. In what follows, we quantize Eqs. (\ref{GP1}) and (\ref{plate1}) to show that strong coupling between the condensate phonons (bogolons) and the flexural modes in the graphene (flexural phonons) is possible, leading to the formation of a sort of phonon-polaritons.

\section{Casimir-Polder interaction with a graphene sheet}

The theoretical approaches to determine the Casimir-Polder energy shift are usually based on second-order perturbation theory \cite{StefanB_Book}. Here, we will discuss how to evaluate the Casimir-Polder potential between a single graphene sheet and a $^{87}$Rb atom within the formalism of macroscopic QED \cite{acta2008}.
To do so, we shall neglecting the possible effects that may arise from the finite size of graphene and assume it to be infinitely extended. For planar structures, the Casimir-Polder potential of an atom in a ground state $| 0 \rangle$ at a distance $d$ away from the macroscopic body can be written as \cite{StefanB_Book}
\begin{equation}
\begin{array}{c}
U_S \pc{d} = \frac{\hbar \mu_{0}}{8 \pi^{2}} \int_{0}^{\infty}
d \xi \xi^{2} \alpha \pc{i \xi} \nonumber \\
\times \int\limits_{0}^{\infty} d k_{\parallel} \frac{e^{-2 k_{\parallel}
\gamma_{0z} d} }{\gamma_{0z}} \left[
\mathrm{R}_{\mathrm{TE}} + \mathrm{R}_{\mathrm{TM}} \left( 1- \frac{2
k_{\parallel}^{2} \gamma_{0z}^{2} c^{2} }{\xi^{2}} \right) \right] ,   
\end{array}
\label{eq:Ucp_1}
\end{equation}
where integration is done along the imaginary frequency axis $\omega= i \xi$ and $k_{\parallel}$ the wave vector in the plane of the interfaces. We have defined $\gamma_{iz} = \sqrt{1+ \varepsilon_{i} (i \xi)\xi^{2}/(c^{2} k_{\parallel}^2)}$, which is the $z$ component of the wave number in a medium with permittivity $\varepsilon_i$ (the index 0 refers to the medium in which the atom is placed). Here, $\alpha(\omega)$ is the atomic polarizability defined for an isotropic atom as
\begin{equation}
\mathbf{\alpha} (\omega) = \lim_{\varepsilon \rightarrow 0} \frac{2}{3 \hbar}
\sum_{k \neq 0} \frac{\omega_{k0} \, |\mathbf{d}_{0 k}|^2}{\omega_{k0}^2-\omega^2-i\omega\varepsilon}\,.
\label{eq:atomicpol}
\end{equation}
The latter is valid in the zero temperature limit and for atoms in the ground state, which will be the case of the present manuscript. All the relevant electromagnetic features of the graphene sheet are cast in the reflection coefficients $\mathrm{R}_{\mathrm{TE}}$ and $\mathrm{R}_{\mathrm{TM}}$ in Eq. \ref{eq:Ucp_1}. In Ref. \cite{PRB84_035446_2011}, the reflection coefficients have been calculate assuming that the dynamics of the quasiparticles in graphene can be described within the $(2+1)-$dimensional Dirac model. Imposing appropriated boundary conditions to the EM field, it is possible to explicitly determine the reflection coefficients. Taking the contribution of the electrons near the Dirac cone, one obtains, for a self-suspend graphene sheet in vacuum,
\begin{eqnarray}
\mathrm{R}_{\mathrm{TM}} &=& \frac{4 \pi \alpha \sqrt{k_{0}^{2} +
k_{\parallel}^{2}} }{4 \pi \alpha \sqrt{k_{0}^{2} + k_{\parallel}^{2}} + 8
\sqrt{k_{0}^{2} + \tilde{v}^{2} k_{\parallel}^{2}}}\,, \\
 \mathrm{R}_{\mathrm{TE}} &=& - \frac{4 \pi \alpha\sqrt{k_{0}^{2} +
\tilde{v}^{2} k_{\parallel}^{2}} }{4 \pi \alpha \sqrt{k_{0}^{2} + \tilde{v}^{2}
k_{\parallel}^{2}} + 8 \sqrt{k_{0}^{2} + k_{\parallel}^{2}}} \,,
\label{eq:reflcoefG}
\end{eqnarray}
where we have defined $k_{0}^{2} = \xi^{2}/c^{2}$ and $\tilde{v}=v_F/c=(300)^{-1}$, with $v_F$ being the Fermi velocity and $c$ the speed of light; $\alpha=1/137$ is the fine structure constant. More elaborated models could be performed, however, for the present conditions, Eq. (\ref{eq:reflcoefG}) provides a very good approximation \cite{PRA88_052521_2013}.\par
There are two regimes according to the atom-surface distances: the near-field, nonretarded limit, and the far-field (retarded) limit. In the nonretarded limit, $|\sqrt{\varepsilon (\omega)}|  \omega d / c \ll 1$, and for the usual Fresnel reflection coefficients, successive approximations yield $U_S \simeq C_3 / d^3$, and we obtain a fitting to Eq. (\ref{eq:Ucp_1}) with $C_3 = -215.65 $~Hz$\mu$m$^3$. The CP potential for a ground-state atom is attractive for a non-magnetic medium \cite{acta2008}. We notice that the sign of the CP potential of a ground-state atom with a conducting plate can be easily understood with an image-dipole model. However, if a resonant coupling between the atomic dipole and a surface excitation occurs, it is possible to obtain a repulsive force \cite{PRL83_5467_1999}.\par
The retarded limit corresponds to the situation where the atom-surface distance $d$ is large when compared to the effective transition wavelength. In this situation, one finds the approximation  $U_S = C_4 / d^4$ also to be valid, with $C_4 = -14.26 $~Hz$\mu$m$^4$. The dependence of the CP potential on the separation distance $d$ as resulting from the numerical integration of Eq. (\ref{eq:reflcoefG}) is depicted in  Fig. (\ref{fig2}). In the following, we will operate in the distance range $d\gtrsim 1~\mu$m, deep in the limit $U_C\sim 1/d^4$. Such a choice is a consequence of three major constraints: i) transverse trap size, as the two-dimensional BEC approximation is only valid if the transverse size of the BEC is much smaller than the distance $d$, ii) the de-confinement effect associated to the CP potential, and iii) the heating of the condensate for small distances. In the following, we assume condition i) to be satisfied (which is true for $\omega_z\sim 2\pi \times 1000$ Hz) and discuss the limitations imposed by ii) and iii) in Sec. V.

\section{Phonon-polariton quantization and avoided crossing}

In order to proceed to a microscopic description of the coupling between the bogolons and the flexural phonons, we quantize the theory. The total Hamiltonian of the system can be defined as $\hat H=\hat H_B +\hat H_C+\hat H_{\rm int}$, where
\begin{equation}
\hat H_B=\int d\mathbf{x} ~\hat \psi^\dagger (\mathbf{x})\left[ -\frac{\hbar^2\nabla^2}{2m}-\mu+g\hat \psi^\dagger (\mathbf{x}) \hat \psi  (\mathbf{x})\right]\hat \psi (\mathbf{x}),
\label{HB}
\end{equation}
is the condensate Hamiltonian, which can be solved in the Fourier basis with $\hat \psi(\mathbf{r})=\sum _\mathbf{k}\varphi_\mathbf{k}(\mathbf{r})\hat a_k$ and $\varphi_\mathbf{k}=\mathcal{V}^{-1/2}e^{i\mathbf{k}\cdot \mathbf{r}}$. Here, $\hat a_k$ represent the bosonic annihilation operator satisfying the canonical commutation relation $\left[\hat a_\mathbf{k},\hat a^\dagger_{\mathbf{k}'}\right]=\delta_{\mathbf{k}\mathbf{k}'}$. Using the Bogoliubov approximation $\hat a_{\bf k}\simeq \sqrt{N_0}+\hat a'_{\bf k}$, with $\hat a'_{\bf k} $ standing for the $k\neq 0$ fluctuations, the condensate Hamiltonian can be simply given by (expressing the summation in terms of the absolute value $k$ only)
\begin{equation}
\begin{array}{ccl}
\hat H_B&=&E_0+\displaystyle{\frac{1}{2}\sum_{k\neq 0} \left[(\epsilon_k+\mu) \left(\hat a_k^\dagger \hat a_k+\hat a_{-k}^\dagger \hat a_{-k}\right)
 \right.}\\
&+& \displaystyle{\mu \left.\left( \hat a_k^\dagger \hat a_{-k}^\dagger+\hat a_{k} \hat a_{-k} \right)\right] }.
\end{array}
\label{HB2}
\end{equation} 
Here, $\epsilon_k=\hbar^2k^2/2m$, $\mu=gn_0$ is the chemical potential and $E_0=N\mu/2$ is the condensate zero-point energy. Similarly, the Hamiltonian for the flexural modes easily follows from the canonical quantization of Eq. (\ref{plate1})
\begin{equation}
\hat H_C=\frac{1}{2}\int d{\bf r} \left[ \left(\rho \partial _t \hat {\bm \eta}(\mathbf{r})\right)^2+D\left(\nabla^2 \hat {\bm \eta}(\mathbf{r})\right)^2+\kappa {\bm \eta}^\dagger(\mathbf{r}) \hat {\bm \eta}(\mathbf{r})\right].
\end{equation}
Expressing the phonon operator in the form 
\begin{equation}
\hat {\bm \eta}(\mathbf{r})=\frac{1}{\sqrt{2}} \sum_{\bf k,\sigma} \varphi_{\bf k}(\mathbf{r})h_{\bf k}(\mathbf{r})\mathbf{e}_\sigma \left(\hat c_{\bf k,\sigma}+\hat c_{\bf k,\sigma}^\dagger \right), 
\end{equation}
with two polarizations $\sigma=(x,y)$ and satisfying the normalisation condition $\langle h_{\bf k}, h_{\bf k'}\rangle=\hbar/(M\omega_C)\delta _{\bf k k'}$, with $M=\rho A$ being the membrane mass, such that the phonon operators obey the bosonic commutation relation $\left[\hat c_{\mathbf{k},\sigma},\hat c^\dagger_{\mathbf{k}',\sigma'}\right]=\delta_{\mathbf{k}\mathbf{k}'}\delta_{\sigma, \sigma'}$, the graphene Hamiltonian reads
\begin{equation}
\hat H_C=\sum_{k ,\sigma}\hbar \omega_C\hat c_{k,\sigma} ^\dagger \hat c_{k,\sigma} .
\end{equation} 
The quantization of the flexural modes is relevant for temperature scales given of the cryogenically cooled atom chip, on which the graphene sheet is supposed to be suspended. Typical experimental conditions involving a dilution refrigerator allow to cool down a carbon nanotube down to 4 K \cite{weiss}. Additionally, according to the recent experiments of Ref. \cite{song}, optomechanical cooling can be used to bring the graphene down to 50 mK, which corresponds to roughly 40 flexural phonon quanta. \par
The interaction Hamiltonian is given by a convolution
\begin{equation}
\hat H_{\rm int}=\int \frac{d\mathbf{x}}{\mathcal{V}}\int d\mathbf{x}'~\hat n (\mathbf{x}')U(\mathbf{x}-\hat {\bm \eta} (\mathbf{x}')).
\end{equation}
Assuming small deformations, we can Taylor expand the potential to first order in $\hat {\bm \eta}$,  $U(\mathbf{x}-\hat {\bm \eta} (\mathbf{x}'))\simeq U(\mathbf{x}-\mathbf{x'})+\bm\nabla U(\mathbf{x}-\mathbf{x'})\cdot \hat {\bm \eta} (\mathbf{x}')$, which yields
\begin{equation}
\hat H_{\rm int}=\sum_k U_k\hat \rho_k+i \sum_k U_k \hat \rho_k{\bm k}\cdot \hat {\bm \eta}_k, \label{Hint}
\end{equation}
with $U_k=\pi C_4 k n_0K_1(kd)/d$ being the Fourier transform of the potential and $\hat \rho_k=\sum_q \hat a_{k+q}^\dagger \hat a_q$ the density operator. The first term can be separated into the condensate and the fluctuation contributions, leading to a change in free-particle energy
\begin{equation}
\sum_k U_k\hat \rho_k=\sum_q U_0\hat a_q^\dagger \hat a_q+\sum_{k\neq 0, q}\hat a_{k+q}^\dagger \hat a_qU_k.
\label{separation}
\end{equation}
The second term of Eq. (\ref{separation}) is of the order of the condensate depletion $\mathcal {O}(N')$, and therefore we should neglect it. By making the substitutions $\epsilon_k\rightarrow \tilde \epsilon_k\equiv ( \epsilon_k+U_kn_0)$ and $E_0\rightarrow E_0+U_0n_0$ in Eq. (\ref{HB2}), and proceeding to a Bogoliubov-Valatin transformation to the excitation operators in the usual way, $\hat a_k'=u_k \hat b_k-v_k \hat b_k^\dagger$, we obtain the modified condensate Hamiltonian as $\hat H_B=\sum_{k}\hbar \omega_B\hat b_k \hat b_k^\dagger$, where $\hbar\omega_B=\sqrt{\tilde \epsilon_k(\tilde\epsilon_k+\mu)}$ is the renormalized Bogolubov spectrum and the transformation coefficients read
\begin{equation}
u_k=\left(\frac{\tilde \epsilon_k+\mu}{\hbar\tilde\omega_B(k)}+\frac{1}{2}\right)^{1/2}, \quad v_k=\left(\frac{\tilde \epsilon_k+\mu}{\hbar\tilde\omega_B(k)}-\frac{1}{2}\right)^{1/2}.
\end{equation}
\par
\begin{figure}[t!]
\includegraphics[scale=0.44]{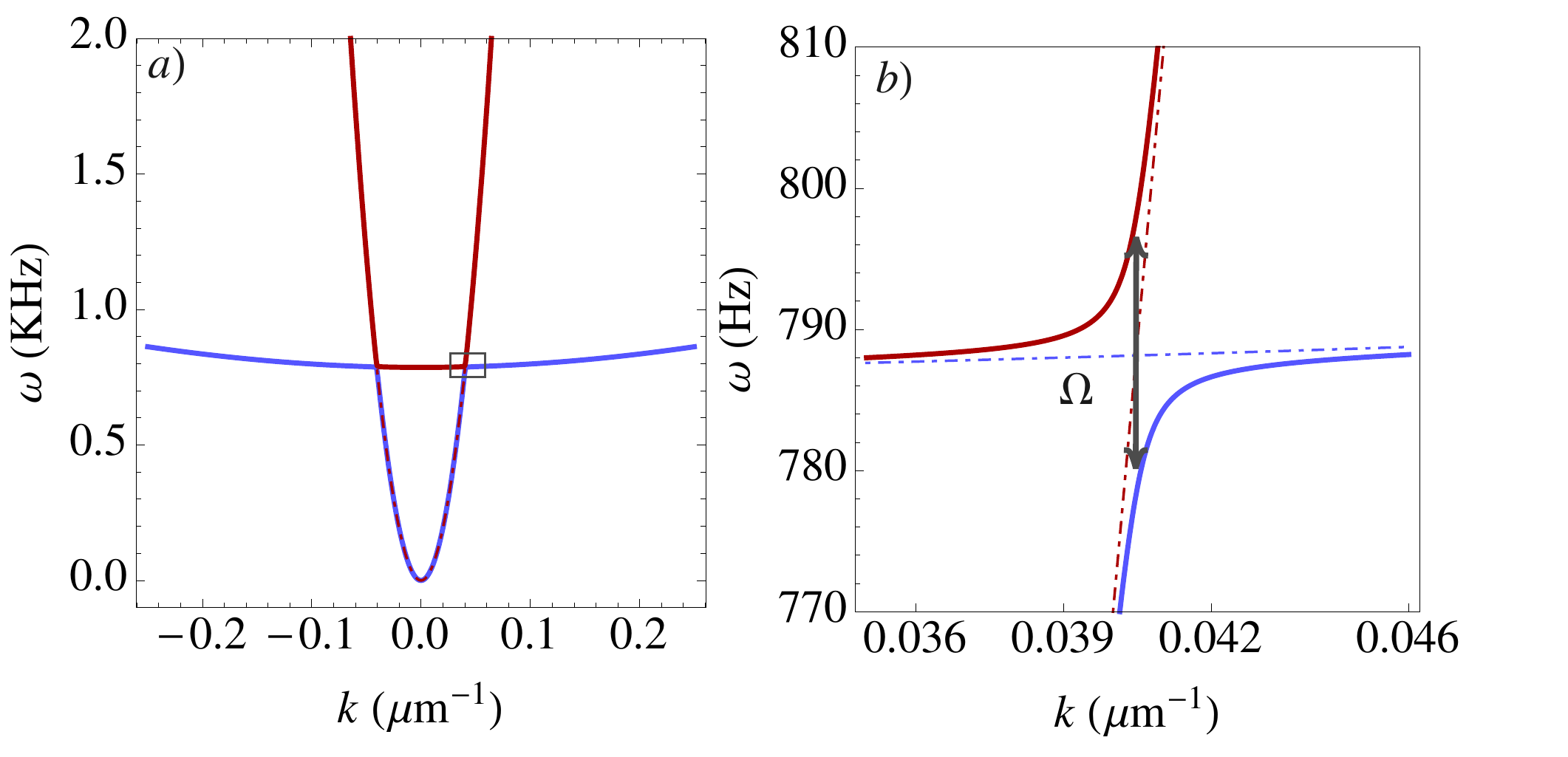}
\caption{(color online) Phonon-polariton dispersion for a system composed of a two-dimensional (2D) condensate of $^{87}$Rb atoms and a graphene sheet interacting via the Casimir-Polder potential $V_C\sim C_4/d^4$. Panels a): Upper (red solid line) and lower (blue solid line) polariton. Panel b): Amplification of the rectangle depicted in panel a). The bare dispersions for the flexural modes (dot-dashed red line) and for the condensate (dot-dashed blue line) are also shown. Both panels are obtained for $d=1.5~\mu$m.}
\label{fig_dispersion}
\end{figure}

The second term of Eq. (\ref{Hint}) can be evaluated by noticing that in the Bogoliubov approximation $ \hat{\rho_k}\simeq \sqrt{N_0}(\hat a_k^\dagger +\hat a_{-k})$, which then yields
\begin{equation}
\hat H_{\rm int}=\sum_{k,\sigma} M_{\kappa,\sigma}\left(\hat b_k^\dagger  \hat c_{k,\sigma}+\hat b_k \hat c_{k,\sigma}^\dagger +\hat b_k^\dagger  \hat c_{k,\sigma}^\dagger +\hat b_k  \hat c_{k,\sigma}\right),
\label{Hint1}
\end{equation}
where $M_{k,\sigma}$ is the $k-$dependent coupling 
\begin{equation}
M_{k,\sigma}=i\sqrt{\frac{N_0}{2}}U_k(u_k-v_k)k h_k \mathbf{e}_k\cdot \mathbf{e}_\sigma.
\end{equation}
The latest two terms in Eq. (\ref{Hint1}) do not conserve the total number of excitations, and therefore will be neglected in the present discussion. This approximation remains valid here as long as the interaction is small compared to the single-particle energy, $\vert M_{k,\sigma}\vert\ll \hbar\omega_B+\hbar \omega_C$, which we will confirm a posteriori. Finally, the total Hamiltonian of the system can be given as 
\begin{equation}
\begin{array}{ccl}
\hat H &\simeq& \displaystyle{\sum_{k}\left(\Delta \hat b_k^\dagger \hat b_k+\sum_{\sigma}\hbar\omega_C\hat c_{k,\sigma}^\dagger \hat c_{k,\sigma} \right. }\\
&+&\left. \sum_\sigma M_{k,\sigma}\hat b_k^\dagger  \hat c_{k,\sigma}+{\rm h.c.}\right),
\end{array}
\end{equation}
where $\Delta=\pi \vert C_4\vert n_0/d^2$ is the interaction-induced energy shift. The later Hamiltonian can be diagonalized using the so-called Hopfield transformations
\begin{equation}
\hat a_k^{(L)}=\chi_{B,k}\hat b_k -\chi_{C,k} \hat c_k, \quad \hat a_k^{(U)}=\chi_{C,k}\hat b_k +\chi_{C,k} \hat c_k,
\label{pol}
\end{equation}
where $\hat c_k=\sum_\sigma \hat c_{k,\sigma}$ and $\hat a_k{(L)}$ and $\hat a_k^{(U)}$ respectively denote the destruction operators for the lower (L) and upper (U) polaritons. In order to be bosonic, the operator in Eq. (\ref{pol}) must satisfy the normalization condition $\vert \chi_{C,k} \vert ^2 +\vert \chi_{B,k} \vert ^2=1$, which allow us to write the Hamiltonian in the decoupled form
\begin{equation}
\hat H=\sum_{P=L,U}\sum _k \hbar \omega_{P}(k)\hat a_k^{(P)\dagger}\hat a_k^{(P)},
\end{equation}
where the eigenenergies are simply given by

\begin{equation}
\hbar \omega_{U,L}(k)=\frac{1}{2}\left[ \Delta +\hbar\omega_C\pm\sqrt{\left( \Delta -\hbar\omega_C\right)^2+4\vert M_k\vert^2}\right],
\end{equation}
with $\vert M_k\vert^2 =\sum_\sigma M_{k,\sigma}^* M_{k,\sigma}$. The two-mode polariton dispersion is depicted in Fig. (\ref{fig_dispersion}). The graphene (condensate) fraction of the LP (UP) mode almost 1 near the bottom of the dispersion $k\sim 0$; on the contrary, for $k\rightarrow \infty$ the LP (UP) mode is essentially Bogoliubov- (graphene-) like. Moreover, the branches repeal each other at the wave vector $k_*=\sqrt{\Delta/\hbar \beta}$. The effective Rabi frequency $\Omega=2\vert M_{k_*}\vert$ measures the strength of the coupling and is equal to $\Omega\simeq 17$ Hz for $d=1.5~\mu$m, being much smaller that the free-energy scale of the system. \par
\begin{figure}
\includegraphics[scale=0.39]{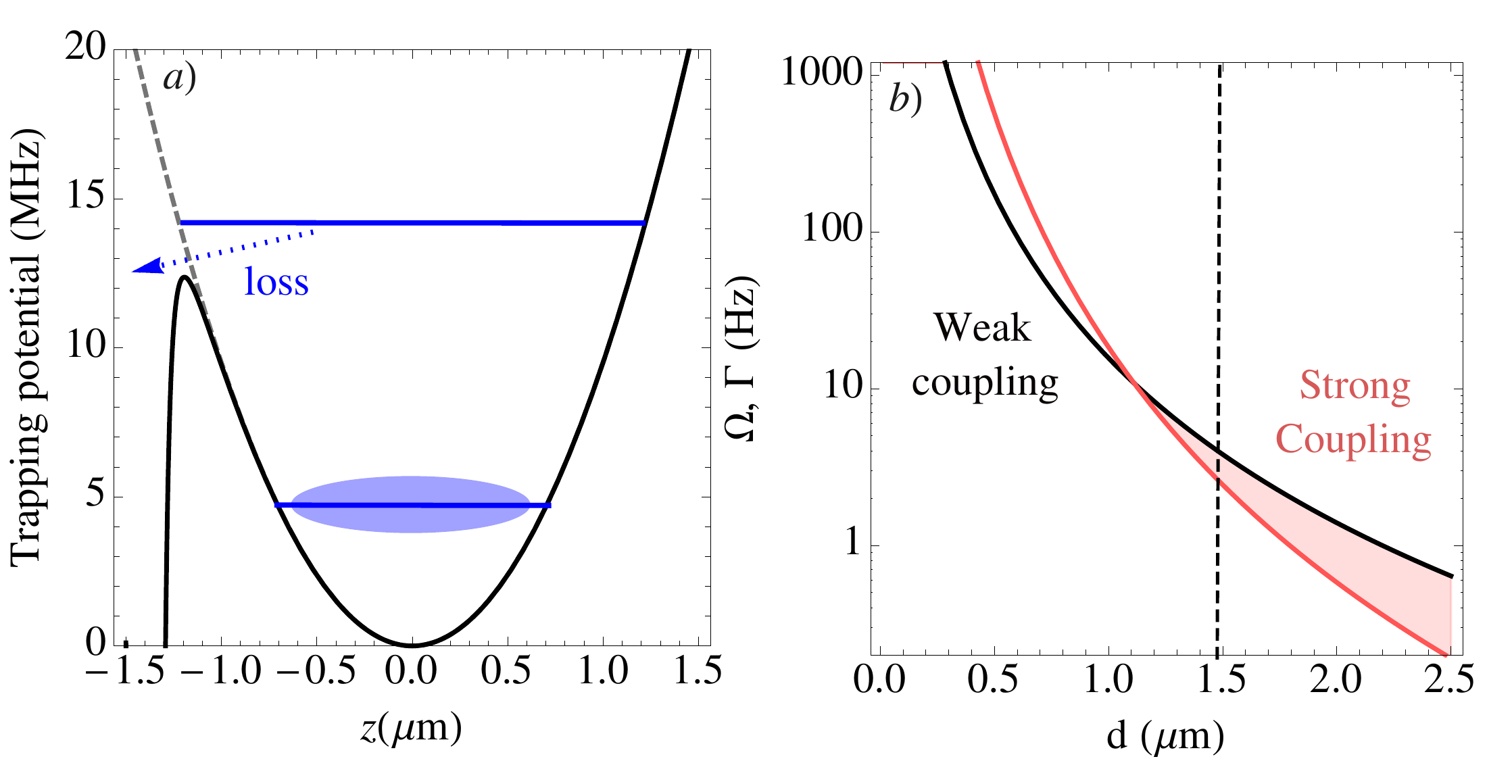}
\caption{(color online) Panel a): Effective trapping potential illustrating the lose trapping effect. If the first transverse excited state lies above the cut-off energy $U_c$, the condensate particles can escape as a consequence of the mechanical heating. Panel b) Effective Rabi frequency (black solid line) and heating rate (red solid line). The shadowed region represents the strong-coupling regime. The dashed vertical line depicts the critical distance $d_c$ (see text). }
\label{fig_coupling}
\end{figure}

In order to demonstrate the conditions for which strong coupling effectively occurs, we must quantify the heating induced in the condensate. As illustrated in Fig. (\ref{fig_coupling} a)), the Casimir-Polder interaction modifies the trapping potential. If $d$ is small enough such that the first excited state of the trap lies above the cut-off energy $U_c=U(z_c)$, where $U(z)=m \omega_z^2z^2/2+C_4/(d-z)^4$ and $z_c\neq 0$ is the solution of the equation $U'(z_c)=0$, the heating may cause the particles to scape the trap \cite{PRL104_143002_2010}. In order to avoid such a situation, we should require the restriction $d>d_c$, where the critical distance is $d_c\simeq z_c+\{2\vert C_4\vert/(3\hbar \omega_z)\}^{1/4}$. For $\omega_z=2\pi \times1500$ Hz, we obtain $d\simeq 1.45~\mu$m. In this situation, the heating rate rules out trap losses and lis related to the creation of in-plane excitations only (condensate depletion), which we calculate with the help of Fermi's Golden Rule
\begin{equation}
\Gamma=\frac{2\pi}{\hbar}\sum_{i,f} \vert \langle i \vert \hat H_{\rm int}\vert f\rangle\vert ^2 \delta\left(\hbar \omega_B-\hbar \omega_C\right),
\end{equation}
where the delta function accounts for the resonant terms only. Here, the initial state $\vert i\rangle=\vert 0_{k_i},1_{\sigma_i,k_i}\rangle$ describes the condensate in the ground state and the graphene phonon with energy $\epsilon_i\simeq \hbar \beta k_i^2$ and polarization $\sigma_i$; the final state $\vert f\rangle=\vert 1_{k_f},1_{\sigma_f,k_f}\rangle$ contains an extra excited state above with flat dispersion $\epsilon_i\simeq \Delta$. By proceeding that way, we capture only the inelastic processes, which after some simple algebra yields the following heating rate
\begin{equation}
\Gamma=\frac{4\pi^2d^2\Delta^3}{M\hbar^2\beta^3}S(k_*)^2K_1^2(k_*d),
\end{equation}
where $S(k)=(u_k-v_k)^{1/2}$ is the BEC static structure factor. The strong coupling regime is achieved for $\Omega\gg \Gamma$, which corresponds to separation distances of the order of 1.5 $\mu$m, deep in the regime where the CP potential considered here is valid (see Fig. (\ref{fig_coupling}b)). The weak coupling situation is also possible for shorter distances, where the heating of the BEC becomes quite appreciable. Due to fast variation of the CP potential, the transition between the strong and weak coupling regimes is quite sensitive. This requires a condensate to be confined at the sub-micro size corresponding to large values of $\omega_z$, therefore safely lying in the regime where trap deconfining becomes less critical. 

\section{Conclusion}

We have demonstrated that the Casimir-Polder due to the interaction of a Bose-Einstein condensate and a graphene sheet can be perturbed by the mechanical out-of-plane vibrations and by the Bogoliubov excitations. At cryogenic temperatures, the Kirchoff-Love flexural modes can be quantized, therefore coupling to the Bogoliubov excitations of the condensate. As a result, a phonon-polariton is formed for sufficiently large separation distances, for which the effective Rabi frequency (the coupling strength) dominates over the heating rate. Our results may motivate a scheme for the sympathetic cooling of monolayer graphene via vacuum-fluctuation forces. In a feasible experimental situation, such a cooling scheme may be implemented in combination with an optomecanical cooling protocol, in order to pre-cool the membrane down to a temperature of few tens of mK \cite{song}. We may think that the sympathetic cooling via the Casimir-Polder interaction may then be used to further cool the zero-point mode, since the condensate temperature is of a few tens of nK. In fact, the major interest in cooling the mechanical modes of graphene lies in the perspective of controlling its transport properties, as the electronic response at low temperatures is intimately related by the scattering with the flexural phonons \cite{eduardo}.

\end{document}